# Steganographic Routing in Multi Agent System Environment

Krzysztof Szczypiorski, Igor Margasiński, Wojciech Mazurczyk

Warsaw University of Technology, Faculty of Electronics and Information
Technology, Institute of Telecommunications 15/19 Nowowiejska Str.
00-665 Warszawa, Poland
{k.szczypiorski, i.margasinski, w.mazurczyk}@tele.pw.edu.pl

**Abstract**: In this paper we present an idea of trusted communication platform for Multi-Agent Systems (MAS) called TrustMAS. Based on analysis of routing protocols suitable for MAS we have designed a new proactive hidden routing. Proposed step-agents discovery procedure, as well as further routes updates and hidden communication, are cryptographically independent. Steganographic exchange can cover heterogeneous and geographically outlying environments using available cross-layer covert channels. Finally we have specified rules that agents have to follow to benefit the TrustMAS distributed router platform.

## 1. Introduction

Decentralization of the operations was recognized as a valid paradigm in the early 1960's. It was confirmed to be robust and efficient and it was used as a fundament in creating architecture for such breaking ideas like Internet and grid computing. Nowadays we are witnessing expansion of distributed systems and services ahead with for example peer-to-peer overlays [1]. Today the most promising application of the distributed operations are agents ([19], [22]). History of the agents tracks back to distributed artificial intelligence (DAI) and distributed problem solving (DPS) concepts [19]. We can define them as the independent software components that are able to act autonomously and which can represent another entity (e.g. human). Moreover systems that consist of many agents interacting with each other form MAS [41].

Combining MAS platforms with steganographic techniques enables secure and unrestricted hidden communication among trusted multi-agent population and it is a novel contribution of the TrustMAS. In our research we were focused on providing trusted communication between chosen agents using steganographic channels. In this way distributed steganographic router is formed.

## 2. Backgrounds

### 2.1 Agents and trust in MAS

Agents can be generally classified as *stationary* or *mobile* agents. The main difference between both types is that stationary agent resides only on a single platform (host that agent operates on) and mobile one is able to migrate from one host to another while preserving its data and state.

Agents can be characterized with the following properties [8]:

- *Interaction* - by performing actions agents influence the environment they operate in,
- *Flexibility* - that can be defined as an ability of the agents to be responsive to the changes occurring in its environment and to interact with other agents (social property) to achieve common goal (proactive property) [22],
- *Autonomy* - no direct intervention of humans (or other entities) is needed for agents to act.

The main advantages of the systems that utilize agents include: fault tolerance (as it is harder for intruder to interrupt communication when it is distributed), scalability and flexibility, performance improvements, lightweight design and an ability to be assigned to different tasks to perform. The most common applications of MAS varies from network monitoring (e.g. IDS/IPS systems [21]) and management, information filtering and gathering (e.g. Google), building self healing, high scalable networks or protection system [42] to transportation, logistics and other (e.g. graphic computer games development [43]).

MAS systems are implemented based on platforms which are the tools to build multi-agent systems. Nowadays the most popular platforms include: JADE [52], AgentBuilder [53], JACK [54], MadKit [55] and Zeus [56]. Such tools simplifies the implementation of multi-agent systems.

Providing security for MAS is crucial as nowadays this technology's global scale development is still limited by security constrains and vulnerabilities ([23], [4], [16], [33], [20]). Classical security model based on central, well secured bastion paradigm is no longer sufficient, because in new distributed network environment agents are ideal attack targets for any malicious operations. Moreover the agents, themselves, are prefect attack tools. The most important attacks that can be performed using MAS include: spamming, DoS (Denial of Service) and spoofing ([16], [21]).

On the other hand mobile agents create dynamic environment and have to be able to establish ad-hoc trust relations to perform intended tasks collectively and effectively. Particularly challenging goals are authentication process where an identity of agent may be unknown and authorization decisions where a policy should accommodate to distributed and changing structure. Trusted cooperation in heterogeneous MAS environment requires not only trust establishment but also monitoring and adjusting existing relations.

Main two concepts of the trust establishment in a distributed environment ([28], [27], [31]) are a *reputation based* ([12], [26], [2], [36], [34]) and a *credential (or rule) based* trust management (TM) ([5], [6], [11], [39], [20], [40]). The first one utilizes information aggregated by system entities to evaluate reputation of chosen entity. Basically, decisions are made according to recommendations from other entities where some of them can be better than others. An example of the reputation computation system can be influential PageRank developed by Google. The second solution – the credential based TM – utilizes secure (e.g. cryptographically signed) statements about a chosen entity. Well known credential based platform is Public Key Infrastructure where role of credentials fulfill X.509 certificates. Essential in reputation evaluation is a presence of a risk factor. These flexible solutions do not exclude wrong decisions. Credential based decisions are more reliable but require well defined semantics.

## 2.2 Steganography

Information hiding techniques such as network, audio, image and text steganography, can became a powerful tool that can be used to establish secure and stealth communication ([18], [25], [38], [32], [3], [24], [35]) among trusted agents. Most of contemporary, widely available implementations of steganographic systems are dedicated to the multimedia applications – hidden data is distributed in sound files, images and movies. A focus on a content exchange in application layer of network model (e.g. watermarking as an intellectual property rights protection tool) can be observed. Steganographic solutions located in network protocols are not relatively widespread, but they exist – most of them rely on usage of communication protocol's optional fields or untypical values from correction codes space.

For MAS environment we propose a distributed steganographic router which will provide ability to create the covert channels between chosen agents. Paths between agents can be built with the use of any of the steganographic methods in any OSI RM layer and be adjusted to the heterogeneous characteristics of a given network. The concept of a steganographic router, as stated earlier, is new in the steganography state of the art and also MAS technology seems to be very accurate to implement such router in this environment. To develop safe and a far-reaching agent communication platform it is required to enhance routing process with anonymity. The first concept of network anonymity was introduced in the seminal paper of Chaum [10]. System Mixnet proposed there has become a foundation of modern anonymity systems. The concept of Mixnet chaining with encryption has been used in a wide range of applications such as E-mail ([30],[13]), Web browsing [18], ISDN [33], and general IP traffic anonymization (Freedom [17], TOR [29]). Other solutions [9] seem to play a less important role or, as Crowds [47], can be considered as simplifications of Mixnet. By means of forwarding traffic for others it is possible to provide agents' untraceability. The origin of collaboration intent in this manner can be hidden from untrusted agents and eavesdroppers.

## 2.3 Routing in TrustMAS

Routing protocols in IP networks are changing, as the networks evolved, from distance-vector (e.g. Routing Information Protocol (RIP), Interior Gateway Routing Protocol (IGRP)), link-state (e.g. Open Shortest Path First (OSPF)) and hybrid (e.g. Enhanced Interior Gateway Routing Protocol (EIGRP)) protocols for wired networks to proactive (e.g. Destination-Sequenced Distance Vector (DSDV), Wireless Routing Protocol (WRP), Global State Routing (GSR), Optimized Link State Routing (OLSR)), reactive (e.g. Ad hoc On-demand Distance Vector (AODV), Dynamic Source Routing (DSR), Light-weight Mobile Routing (LMR)) and hybrid (e.g. Zone Routing Protocol (ZRP), Scalable Location Update Routing Protocol (SLURP), Distributed Dynamic Routing (DDR)) protocols for MANETs [50], [51].

In TrustMAS the most important component that proposed distributed steganographic router must posses is routing protocol. The effective routing protocol is vital for agents' communication and their performance. The routing protocol that will be developed for TrustMAS must take into account all specific features that are not to find in any other routing environment. That includes: providing anonymity with random walk algorithm and usage of steganographic methods. Both those aspects affect performance of the routing convergence. The first one influences updates: due to provide anonymity service they must be periodic. The second one affects links' available bandwidth.

That is why the routing protocol for TrustMAS will be designed from the scratch, will be kept as simple as possible so non of the existing routing protocols for MANETs are applicable. It will be a distance vector proactive algorithm (and will be described in details in section 4).

## 3. Architecture and main components of TrustMAS

### 3.1 Agents in TrustMAS: Steganographic Agents (SAs) and Ordinary Agents (OAs)

Two types of agents are present in TrustMAS platform. There are Ordinary Agents (OAs) that uses this platform to benefit from two security services that it provides: trust and anonymity. The second type of agents are Steganographic Agents (StegAgents, SAs) that use TrustMAS to perform hidden communication. The OAs are not aware of the presence of SAs. And even if malicious agents exist and try to uncover SAs and their communication, there are certain mechanisms available in TrustMAS (described later) to limit potential risk of disclosure. Each StegAgent is characterized with its address and steg-capabilities that describe the steganographic techniques that SA can use to create hidden channel to communicate with other SAs.

In TrustMAS, StegAgents may perform steganographic communication in various ways, especially by using methods in different layers of TCP/IP model. In particular, SAs may utilize other than application layer methods by using specialized middleware enabling steganography through all layers in this model. In some cases there is a possibility to use only application layer steganography i.e. image or audio hiding methods. Hidden communication via middleware in different layers gives opportunity for SAs to establish links outside MAS platform. Examples of techniques in different layers of TCP/IP model that enable covert channels includes:

- Application layer e.g. audio, video, still images, text hiding methods,
- Transport and network layer: protocol (network) steganography,
- Data link layer methods depend on available medium e.g. HICCUPS [38] system can be utilized on WLAN links.

Using such cross-layer steganography has certain advantages as it gives more possibilities of exchanging hidden data and it is harder to uncover. However, building the path with many different steganographic methods may introduce delays, therefore some hidden data methods, known from the state of the art, may be not sufficient to carry network traffic (reminding that in some steganographic applications delay is not best measure, because the best one is just to be hidden).

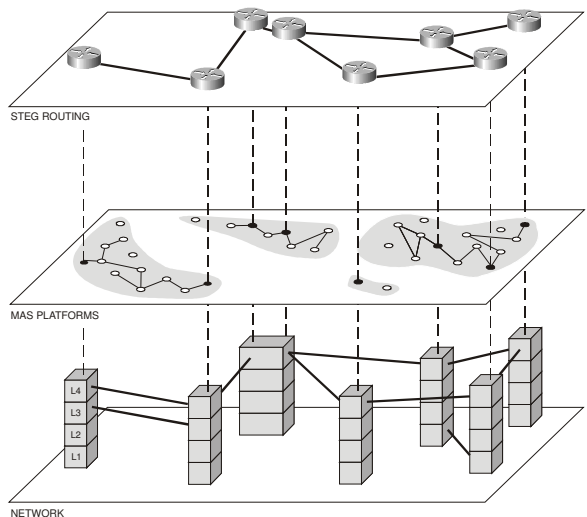

**Figure 1.** Architecture of TrustMAS

### 3.2 TrustMAS architecture

The proposed architecture of TrustMAS can be described on three planes (Fig. 1). In MAS PLATFORMS plane, the gray areas represent homogenous MAS platforms, black dots represent StegAgents and white ones Ordinary Agents involved in TrustMAS. StegAgents act as a distributed steganographic router (Steg-Router) as shown on STEG ROUTING plane. Connections are possible between StegAgents with use of hidden channels, located in various network layers (NETWORK plane), and at the platform level. As mentioned earlier, what steganographic methods will be used to communicate between each StegAgents depend on their steg-capabilities.

The main components that form TrustMAS architecture, that will be described, include: SAs and OAs (section 3.1), trust and anonymity services (section 3.3) and distributed steganographic router (section 3.4).

### 3.3 Security services in TrustMAS: trust and anonymity

Multi-agent systems give opportunity to build an agents' community. In such environments, like in human society, trust and anonymity become important issues as they help agents to build and manage their relationships. That is why we assume that:

- There are no typical behaviors of the agents involved in the particular MAS community,
- All agents may exist and live their lives in their own way - this assumption results in lack of defining agents' interests and gives no information about characteristics of exchanged messages,
- Because our work is focused on information hiding in MAS, we don't assume that any background traffic exists.

These assumptions are rather generic and do help to describe TrustMAS in theoretical way of building steganographic system. In real environment such as IP networks background traffic will exist and will aggravate detecting of the system.

Agents must posses certain level of trust for each other, in order to minimize the uncertainty of the interactions they perform. Moreover agents interactions often have to happen in uncertain, dynamically changing and distributed environment. Trust, as it is usually describing reliability or trustworthiness of the other communication sides, supports agents in making right decisions. When trust value is high the party with which agent is operating gives more chances to succeed e.g. agents need less time to find and achieve their goals. On the contrary, when trust value is low, the choice of the operating party is more difficult, time-consuming and provides less chances for success. In a proposed TrustMAS platform we provide trust and anonymity for each agent that wishes to join it. Main trust model of TrustMAS platform is based on specific behavior of agents – waiting for expected scenario and following dialog process mean that agents are trusted. Other, not included in this work trust models depend mainly on application of TrustMAS and can be changed accordingly.

TrustMAS includes anonymous technique based on random-walk algorithm [46] for providing general purpose anonymous communication for agents. To send a message anonymously the agent sends the message to a randomly chosen agent. The message contains a destination address. Then, the selected agent flips an asymmetric coin to decide whether to forward the message to the next random agent. The coin asymmetry is described by a probability $p_f$. The proxy agent forward the message to the next random proxy agent with the probability $p_f$ and skip forwarding with a probability $1 - p_f$. This probabilistic forwarding assures anonymity because any agent can not conclude if messages received in this manner are originated from their direct sender.

If many agents join TrustMAS it will be easier to hide covert communication exchanged between SAs. All agents that take part in proposed MAS platform benefit from trust and anonymity that is provided for their interactions. But ability of using TrustMAS dictates some conditions: all agents that want to use it are obligated to follow certain rules like e.g. forward discovery steganographic messages according to random-walk algorithm. This is the "cost" that agents have to "pay" in order to benefit from trusted environment.

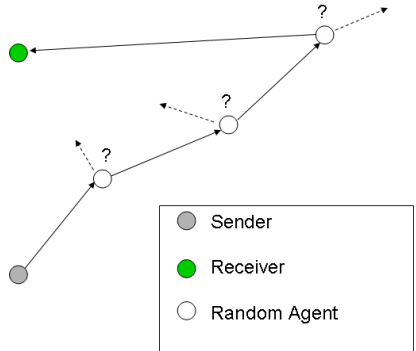

**Figure 2.** Agents Random Walk

### 3.4 Distributed steganographic router (Steg-router)

As described in section 3.2 all the StegAgents in TrustMAS and their ability to exchange information by using hidden channels form distributed steganographic router (Steg-router). Proposed Steg-router is a new concept of building distributed router to carry/convert different covert channels, where typically covert channel is end-to-end connection. Conversion of hidden channels is performed in heterogeneous environment (exp.: hidden information in an image converted into hidden information in WLAN) and the MAS platform is used here as environment to implement this concept. This gives opportunity to evaluate a new communication method and explore new potential threats in MAS environment.

The main component of proposed Steg-router is steganographic routing protocol (Steg-routing protocol) that is described in section 4. It is a distance vector protocol and it uses random walk algorithm (mentioned earlier) to perform discovery of new StegAgents that join TrustMAS (new SAs also perform this algorithm in order to join TrustMAS, to be able to find existing SAs). It also utilizes hello mechanism to build neighbors' relations with other SAs and to detect changes in their presence. We chose a distance vector routing protocol without triggered updates for security reasons - to avoid potential attacks connected with monitoring agents behavior. We can imagine a situation in which the aim of the malicious attack is to observe agents behavior after removing random agent from the TrustMAS. If the removed agent was StegAgent and if the Steg-routing protocol uses triggered updates then suddenly there will be vast activity in the TrustMAS, because triggered updates will be send to announce changes in the network topology. From the same reason distance vector protocol was utilized over the link state or hybrid one. Another drawback of the link state protocol (or hybrid) for our purposes is that it has greater requirements on processing time and memory then distance vector and agents may be lacking in both those aspects.

## 4. Steg-routing protocol

A typical distance vector routing protocol operates generally in the following way: each node sends periodical routing updates (its entire route table) to all their neighbors. That is why proposed steg-routing protocol will be characterized by describing three mechanisms:

- Discovery and maintenance of the neighbors (section 4.1),
- Exchanging routing tables (section 4.2),
- Creating steg-links and steg-paths (section 4.3).

### 4.1 Discovery of new SAs and maintaining neighbors table

As stated above, all the agents involved in TrustMAS perform anonymous exchange based on random-walk algorithm. In this procedure each agent uses asymmetric coin to decide if it passes data or not to randomly chosen agent (StegAgents or other involved in TrustMAS). StegAgents uses this procedure to send anonymous message with embedded stegmessage that consists of:

- StegAgent's address,
- StegAgent's steg-capabilities (available steganographic methods to use for covert communication).

Such mechanism is analogous to sending hello packets to the neighbors in classical distance vector protocols, where it is responsible for discovery and maintenance of the neighbors table. In proposed protocol random walk algorithm performs only discovery role. So the discovery phase is performed by SAs that are already involved in TrustMAS and by new SAs that want to join it.

Moreover, each StegAgent will maintain two tables: neighbors and routing table. Neighbors table is created based on the information obtained from random-walk algorithm operations. The neighbor relation is formed between two StegAgents if there is a steg-link that connects them. Maintenance of the actual information in neighbors table is achieved by sending, periodically, hello packets through formed steg-links (covert channels – connection using steganography to next hop SA). Such a solution helps to identify the situation when one of the StegAgents becomes unavailable.

So the discovery and maintenance procedure from the new StegAgent point of view, that wishes to join TrustMAS, can be described in the following steps:

- Each SA (joining or already involved) uses random walk algorithm to discover other SAs in TrustMAS. Fig. 3 presents the situation, for the case, when new SA tries to connect to existing, already interconnected StegAgents,
- Each agent (SA or OA) passes or drops the discovery stegmessage sent based on the random walk algorithm. In this way new or existing SAs are learned,
- Based on the information collected from the first two steps, steg-links are formed between new StegAgent and found ones if their steg-capabilities match,
- Two SAs become neighbors if the steg-link exists between them. The corresponding entry is added in new SA neighbors table,
- Each SA sends periodically hello packets through available steg-links (Fig. 4) to check if the neighbor is still

available,
- If a hello packet is received by SA it refreshes corresponding entry in its neighbors table. If the hello packet is not received during set period of time it is removed from the neighbors table.

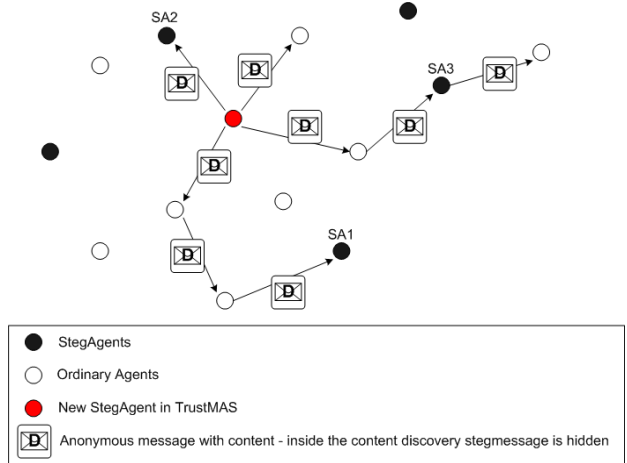

**Figure 3.** Discovery mechanism with random walk algorithm in TrustMAS for new SA

Outside the platforms connections are learned from fixed relations. Collected information helps to form routing tables.

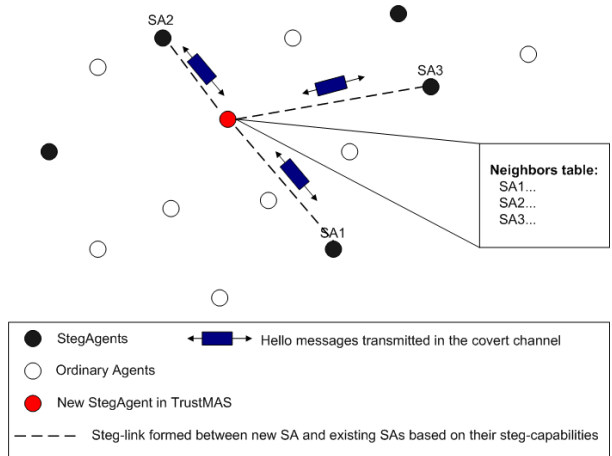

**Figure 4.** Forming steg-links between StegAgents and creating neighbors table

### 4.2 Exchanging routing information

TrustMAS uses steganographic channels to exchange routing tables between StegAgents. These routing updates are sent also at regular intervals to finally achieve proactive hidden routing. Routing proactivity provides unlinkability of the steganographic connections and discovery process. This procedure as well as further hidden communication is cryptographically independent.

To show how the routing information is exchanged we will continue the scenario from the section 4.1, where the new SA joins TrustMAS. After the discovery phase, when the new SA's neighbors table possesses actual information it receives entire routing tables from its neighboring StegAgents (Fig. 5).

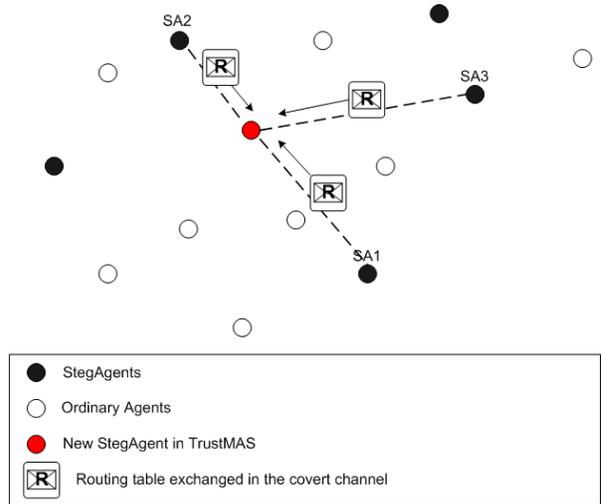

**Figure 5.** Exchanging routing information between SAs

Then the routing information is exchanged periodically between SAs. When new SA receives the routing tables from its neighbors it is able to learn about other distance SAs and how to reach them. Based on this information it can also form new steg-links with other SAs (Fig. 6).

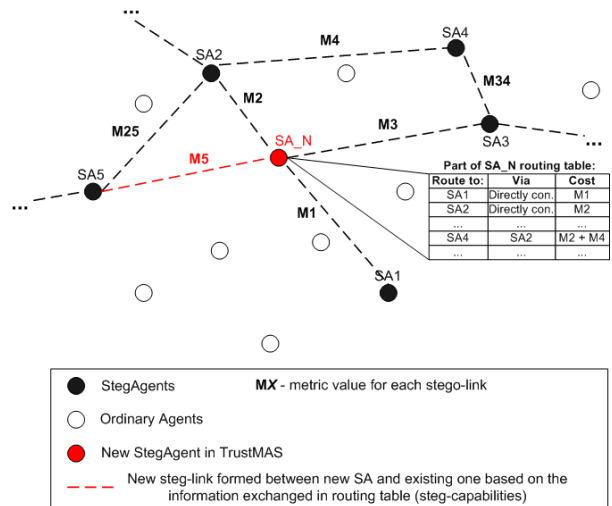

**Figure 6.** New StegAgent learns about other SAs n TrustMAS

If one of the SAs becomes unavailable, the change is detected with the hello mechanism. Then routing table is updated and the change is sent to all the neighbors in the neighbors table, when there is time (periodic) to send the entire routing table.

Each routing entry in the routing table represents best available steg-path to distance StegAgent with its metric. The metric is based on:

- Available capacity of the steg-links along the end-to-end steg-path,
- Introduced delays along the steg-path,
- Available steganographic methods – for security reasons some steganography methods may be preferred then others (e.g. because they are more immune to steganalisys).

The algorithm of StegAgent hidden routing protocol (Steg-routing) can be also expressed in the following pseudo code:

**Algorithm 1**

```
(1)    randomWalkRequest ← listenMAS()
(2)    routingUpdateRequest ← listenNETWORK()
(3)    hello ← listenNETWORK()
(4)    do
(5)    {
(6)      if (randomWalkPeriod + random(fluctuationRW)
(7)    exceeded)
(8)        sendRandomWalk(myAddress, myCovertChannels)
(9)      if (routingUpdatePeriod + random(fluctuationRU)
(10)   exceeded)
(11)       sendRoutingUpdate(myRoutingTable)
(12)     if (helloPeriod + random(fluctuationH) exceeded)
(13)       sendHello(myNeighboursTable)
(14)     if (randomWalkRequest)
(15)     {
(16)       if (findStegMsg(randomWalkRequest))
(17)       {
(18)         foundAddress, foundCovertChannels ←
(19)   uncover(randomWalkRequest)
(20)         if (isNewEntry(foundAddress,
(21)   foundCovertChannels))
(22)         {
(23)           myRoutingTable ←
(24)             updateMyRoutes(foundAddress,
(25)   foundCovertChannels)
(26)           sendRoutingUpdate(myRoutingTable)
(27)         }
(28)       }
(29)       forwardRandomWalk(randomWalkRequest)
(30)     }
(31)     if (routingUpdateRequest and
(32)   findChanges(routingUpdateRequest))
(33)     {
(34)       myRoutingTable ←
(35)         updateMyRoutes(routingUpdateRequest)
(36)       sendRoutingUpdate(myRoutingTable)
(37)     }
(38)     if (hello)
(39)     {
(40)       myRoutingTable ←
(41)         updateNeighborLastHelloTime(hello)
(42)     }
(43)     for each neighbor ← entry(myNeighborTable)
(44)       if(helloTimeout(neighbor) exceeded)
(45)       {
(46)         myNeighborTable ←
(47)           removeEntry(neighbor)
(48)         sendRoutingUpdate(myRoutingTable)
(49)       }
(50)   }while (∞)
(51)
(52)   subroutine sendRandomWalk(address, channels)
(53)   {
(54)     destination ← selectRandomAgent(myPlatform)
(55)     sendViaMAS(destination, cover(address,
(56)   channels))
(57)   }
(58)
(59)   subroutine forwardRandomWalk(message)
(60)   {
(61)     if (coinFlip(pf) = heads)
(62)     {
(63)       destination ← selectRandomAgent(myPlatform)
(64)       sendViaMAS(destination, message)
(65)     }
(66)   }
(67)
(68)   subroutine sendRoutingUpdate(table)
(69)   {
(70)     for each destination ← entry(myNeighborTable)
(71)       sendViaNETWORK(destination, cover(table))
(72)   }
```

### 4.3 Additional improvements to limit convergence time

If the steg-routing protocol operates like described in sections 4.1 and 4.2 the following scenario may occur: new StegAgent uses random walk algorithm and discovers two existing SAs (SA1 and SA2 - Fig. 7).

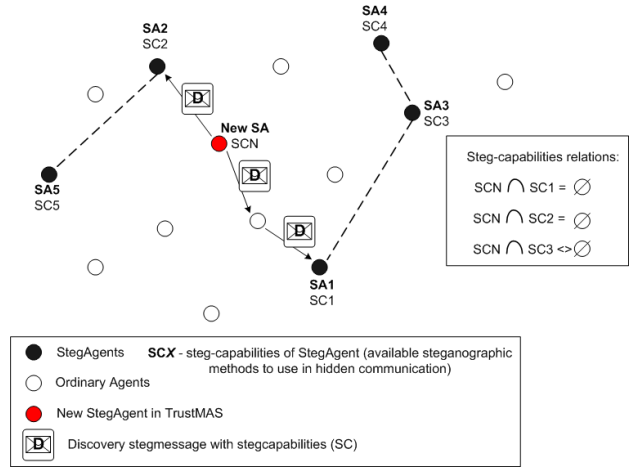

**Figure 7.** Scenario: new StegAgent discovers two SAs but can not communicate with them due to steg-capabilities incompatibility

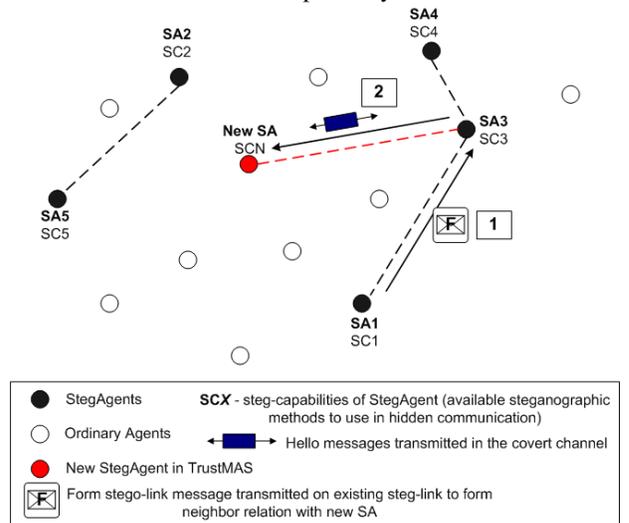

**Figure 8.** Mechanism for improving convergence that enables other SAs to form a steg-link with new SA that shares the same steg-capabilities

Unfortunately both StegAgents: SA1 and SA2 posses steg-capabilities that are not compatible with new StegAgent ones. That means that it is unable to exchange either hello packets nor routing tables, because the steg-links are not formed. In this case the following mechanism can be utilized to improve convergence as showed in Fig 8.

The main idea of this mechanism is as follows: if existing SA is discovered by new SA that was not yet known and their steg-capabilities are incompatible (like New SA and SA1 in Fig. 8) it sends "form steg-link" message (marked as 1, F message in Fig. 8) to one of its neighbors that shares the same steganographic methods as new SA (it can choose this neighbor by inspecting its routing table where the steg-capabilities are also stored). Such message (form steg-link) must contain an address of new SA (SA_address) and its steg-capabilities (SA_steg_capabilities) and can be formed as:

**form steg-link SA_address SA_steg_capabilities**

In Fig. 8 SA1 sends "form steg-link" message to SA3, because it knows that SA3 possess compatible steg-capabilities with joining SA. When StegAgent SA3 receives this message it send the hello packet to the new SA to form a neighbor relation and to form a steg-link. Then the routing table is exchanged and new SA learns about other SAs in the TrustMAS.

### 4.4 Creating steg-links and steg-paths

The end-to-end connection between two distant StegAgents is called steg-path. Every steg-path is created based on available steg-links. The algorithm of forming a steg-path uses metrics that are set for each steg-link. Routing metrics in TrustMAS are calculated based on the steganography methods, its capacity and introduced delays.

If two hops are available, the steg-link is chosen to the path, if it possesses higher capacity value, introduce less delay and uses more preferred steganographic method. Also there is a situation possible that on one steg-link two or more steganographic methods are available. In this case metrics are calculated for each steganographic method and the best is chosen (Fig. 9).

Each SA is also responsible, if it is necessary, for converting steganographic channels according to the next hop SA steg-capabilities. In this way a steganographic router functionality is provided.

The algorithm that each StegAgent uses to choose steg-path can be expressed in the following pseudo code:

**Algorithm 2**

```
(1)    if (newDataToSend)
(2)    {
(3)        paths ← findPathsMatch(myRoutingTable,
(4)    destination)
(5)        if (count(paths) > 1)
(6)        {
(7)            calcMetricsForPaths(paths, capacity, delay,
(8)    steg_method)
(9)            BPath ← chooseBestPath(paths)
(10)           sendData(BPath)
(11)       }
(12)       else
(13)           if (count(paths) = 1) sendData(paths)
(14)           else noPathFound()
(15)   }
```

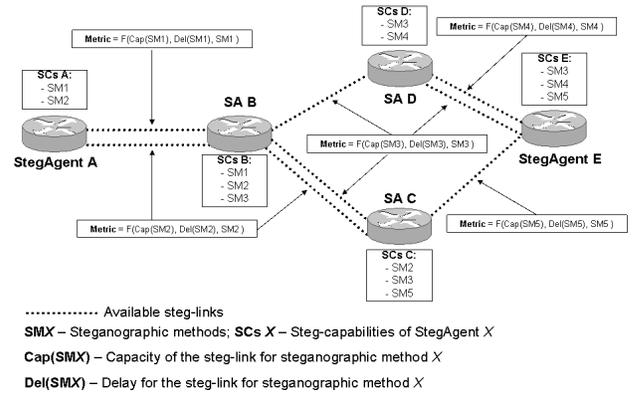

**SMX** – Steganographic methods; **SCs X** – Steg-capabilities of StegAgent X
**Cap(SMX)** – Capacity of the steg-link for steganographic method X
**Del(SMX)** – Delay for the steg-link for steganographic method X

**Figure 9.** Example of SAs with their available steg-links and calculated metrics

Created and maintained routing table enables StegAgent to send data via hidden channels, where metrics are calculated based on the available steganographic methods.

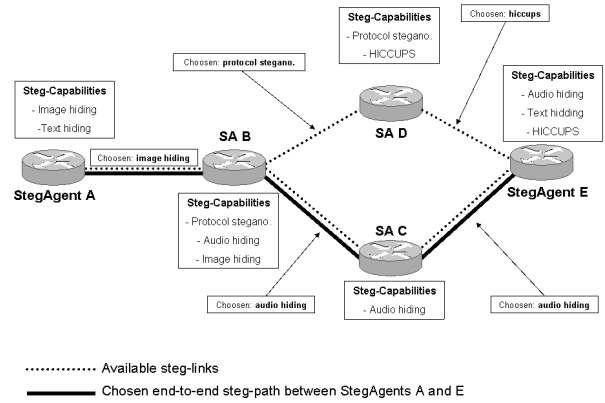

········ Available steg-links
▬▬▬▬ Chosen end-to-end steg-path between StegAgents A and E

**Figure 10.** Example of forming steg-path based on available steg-links

Fig. 10 shows how an example end-to-end steg-path is formed based on exemplary steganographic methods. As mentioned earlier each steg-path consist of certain number of the steg-links (connection to next hop SA; steg-link is e.g. between SA A and SA B in Fig. 10).

## 5. Conclusion and future work

We have presented concept of a distributed steganographic router that provides ability to create the covert channels between chosen agents. Paths between agents can be built with the use of any of the steganographic methods in any network layer and be adjusted to the heterogeneous characteristics of a given network.

Future work will cover performance analysis of the proposed steganographic routing, its convergence time, available range, and potential limitations.

### Acknowledgment

This material is based upon work supported by the European Research Office of the US Army under Contract No.

## Authors' Biographies

**Krzysztof Szczypiorski** received M.Sc. (1997) and Ph.D. (2007) in telecommunications both with honors from Faculty of Electronics and Information Technology, Warsaw University of Technology (WUT, Poland); assistant professor at WUT; main research interests: network security and steganography, wireless networks, privacy in virtual society; author of over 40 scientific papers and over 30 invited talks on information security, telecommunications and electronic commerce; leader of Network Security Group (secgroup.pl).

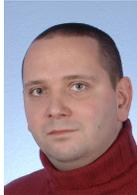

**Igor Margasiński** is a research assistant at Institute of Telecommunications, Faculty of Electronics and Information Technology, Warsaw University of Technology. Received B.sc. (2002) and M.Sc. (2003) in telecommunications at WUT; research interests: network security, privacy enhancing technologies, anonymous networks – in particular peer-to-peer overlays and mobile agent systems, anonymity modeling and metrics, and traffic performance modeling for anonymous systems.

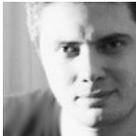

**Wojciech Mazurczyk** received the B.sc. (2003) and M.Sc. (2004) in telecommunication both from Faculty of Electronics and Information, WUT; research assistant at WUT; he is now pursuing a PhD degree in network security; main research interests: information hiding techniques, network security and multimedia services; member of Network Security Group (secgroup.pl).

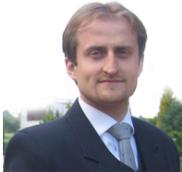